\documentclass{emulateapj}

\newcommand{\invdetsp}{\frac{1}{\sqrt \gamma}}
\newcommand{\invdetst}{\frac{\Lambda}{\alpha \sqrt \gamma}}

\shorttitle{Energy Extraction from Rotating Black Holes}
\shortauthors{Menon \& Dermer}

\received{}
\begin{document}

\title{Analytic Solutions to the Constraint Equation for a Force-Free
Magnetosphere around a Kerr Black Hole}

\author{Govind Menon\altaffilmark{1,2,3} \& Charles D. Dermer\altaffilmark{2}}
%\affil{E. O. Hulburt Center for Space Research, Code 7653,\\
%Naval Research Laboratory, Washington, DC 20375-5352}
\altaffiltext{1}{Troy University, Troy, Alabama, 36082}
\altaffiltext{2}{E. O. Hulburt Center for Space Research, Code 7653,
Naval Research Laboratory, Washington, DC 20375-5352}
\altaffiltext{3}{On sabbatical leave from Troy University.}

%\email{gmenon@troy.edu}

%\author{Charles D. Dermer}
%\affil{E. O. Hulburt Center for Space Research, Code 7653,\\
%Naval Research Laboratory, Washington, DC 20375-5352}
%\email{dermer@gamma.nrl.navy.mil}

\begin{abstract}
The Blandford-Znajek constraint equation for a
stationary, axisymmetric black-hole force-free magnetosphere
is cast in a 3+1 absolute space and time formulation, 
following Komissarov (2004). We derive an 
analytic solution  
for fields and currents to the constraint equation in 
the far-field limit that satisfies the Znajek condition 
at the event horizon. This solution generalizes the Blandford-Znajek
monopole solution for a slowly rotating black hole to black holes 
with arbitrary angular momentum. Energy and angular momentum extraction
through this solution occurs mostly along the equatorial plane. 
We also present a nonphysical, reverse jet-like solution.
\end{abstract}


\section{Introduction }

\citet{pen69} recognized the possibility to extract the spin energy 
of a black hole using particle decay in negative energy orbits
within the ergosphere. Based on studies
of force-free pulsar magnetospheres, \citet{bz77} proposed that
rotational energy could be extracted through currents 
flowing in the black hole's magnetosphere. In this picture, 
strong electric and magnetic fields are induced by gravito-MHD (GMHD)
processes.  \citet{bz77} derived the equations for a stationary,
axisymmetric force-free magnetosphere in curved spacetime, and 
reduced the set of equations to a central constraint equation 
relating toroidal magnetic field $H_\varphi$ to the charge 
density $\rho$ and toroidal current density $J_\varphi$. They
also found a perturbative solution to the constraint equation
valid in the limit $a/M \ll 1$, where
$M$ is the black hole mass and $a$ is the angular momentum per unit mass.

\citet{tmd82} and \citet{mdt82} developed this theory in 
a more intuitive ``3+1" formulation that led to 
the membrane paradigm \citep{tpm86}, where
the equations of GMHD were written using the familiar electric 
and magnetic 3-vectors in absolute space whose time dependence
is governed by Maxwell-type equations. \citet{kom04} recently
presented the essential equations of this formulation 
in a form useful for numerical studies, and helped resolve
questions \citep{pc90,pun01} relating to energy extraction 
in the membrane paradigm. 

The equations presented by \citet{kom04}
provide a useful starting point to search 
for analytic solutions. Here we use these equations
to rederive the constraint equation of \citet{bz77}
in the 3+1 form. This brings forth a clear understanding
of the nature of the poloidal functions defining the 
currents and fields. We have discovered an analytical solution
valid for arbitrary angular momentum that reduces to 
the monopole solution of \citet{bz77} in the limit 
of $a/M \ll 1$. This solution, which satisfies the 
\citet{zna77} regularity condition,
permits energy extraction preferentially along the 
equatorial direction of the Kerr black hole. 

The modified Maxwell's equations
in curved space-time are given in Section 2, and
the equations for a force-free magnetosphere are presented
in Section 3.  In Section 4, we construct the form of 
fields and currents for a given poloidal function $\Omega$.
The governing constraint equation for this function 
is given in Section 5.
Solutions to this equation are derived in Section 6, and
we summarize in Section 7.

\section{Electrodynamics In Absolute Space}

While Maxwell's equations preserve all of its elegance in a covariant
formalism on a four dimensional manifold, it distracts
from some of the simple (far-field) solutions that it might permit. 
With this is mind, we briefly state the
essential equations of electrodynamics in an absolute three dimensional space.
The recent paper by \citet{kom04} explains how these equations are
derived. 

The construction of absolute space is facilitated by noting that an
arbitrary spacetime metric can be written in the form 

\begin{equation}
ds^2=(\beta^2-\alpha^2)dt^2+2\beta_i dx^i dt + \gamma_{ij} dx^i dx^j .
\label{ds2}
\end{equation}

The functions ${x^i}$ serve as coordinates for our spacelike hypersurfaces
defined by constant values of $t$. Consider one such hyperspace $\Sigma$
defined by the region $t=0$. We can think of electric and magnetic fields
($E$ and $B$) as objects existing in our absolute space $\Sigma$. The time
evolution equations for $E$ and $B$ in the presence of a charge density
$\rho$ and electric current density vector $J$ in our absolute (curved)
space endowed with a metric $\gamma_{ij}$ are given by the following set of
Maxwell's equations:
\begin{equation}
\nabla\cdot B=0,
\label{max1}
\end{equation}

\begin{equation}
\partial_tB + \nabla \times E = 0,
\label{max2}
\end{equation}
and their inhomogeneous counterparts,

\begin{equation}
\nabla\cdot D=\rho,
\label{max3}
\end{equation}

\begin{equation}
-\partial_tD + \nabla \times H = J.
\label{max4}
\end{equation}
It is important to remember that $E$, $B$, $D$, $H$, and $J$ are vectors in our
three dimensional absolute space $\Sigma$, and in general are time dependent. Also,
$\nabla$ is the covariant derivative induced by the metric $\gamma_{ij}$ on $\Sigma$. 
As usual, the curl of a vector field is defined by the expression

\begin{equation}
(\nabla \times A)^i = \epsilon^{ijk} \nabla_j A_k,
\label{curl}
\end{equation}
where $\epsilon^{ijk}$ is the completely antisymmetric pseudotensor such that
$\epsilon^{123} = \frac{1}{\sqrt\gamma}$, and $\gamma = {\rm Det}(\gamma_{ij})$.
It is easily seen that Maxwell's equations imply the continuity equation

\begin{equation}
\partial_t\rho + \nabla \cdot J = 0.
\label{cont}
\end{equation}
Unlike its flat space counterparts, even in regions of negligible electric
and magnetic susceptibilities, $E \neq D$, and $B \neq H$. Indeed, it can be shown that
they instead satisfy the consitutive relations

\begin{equation}
E = \alpha D + \beta \times B,
\label{suscept1}
\end{equation}
and
\begin{equation}
H = \alpha B - \beta \times D.
\label{suscept2}
\end{equation}

Of interest are
spacetimes admitting Killing fields corresponding to axial
symmetry $(m)$ and stationarity. Consequently, Noether's theorem
imply energy and angular momentum conservation laws. They can be stated
in the form

\begin{equation}
\partial_te + \nabla \cdot S = -(E\cdot J),
\label{cons1}
\end{equation}
and
\begin{equation}
\partial_tl + \nabla \cdot L = -(\rho E + J\times B)\cdot m.
\label{cons2}
\end{equation}
Here,
\begin{equation}
e = \frac{1}{2}(E\cdot D + B \cdot H)
\label{energy}
\end{equation}
is the volume density of energy, and
\begin{equation}
l = (D \times B)\cdot m
\label{ang}
\end{equation}
is the density of angular momentum,
\begin{equation}
S = E\times H
\label{engflux}
\end{equation}
is the flux of energy, and
\begin{equation}
L = -(E\cdot m)D - (H\cdot m)B + \frac{1}{2}(E\cdot D + B \cdot H)m
\label{angflux}
\end{equation}
is the flux of angular momentum.

\section{Stationary, Axisymmetric Force Free Magnetospheres}

The condition that the magnetosphere is force free brings about enough
structure into Maxwell's equations to enable the introduction of a
streaming function that will help us visualize the field structure in
geometric terms. It is traditional to use spheroidal spatial coordinates
given by ${x^i} = (r, \theta, \varphi)$ such that $m=\partial_\varphi$.
Assumptions of stationarity and axissymmetry imply that
$\partial_\varphi g_{\mu \nu}=0=\partial_t g_{\mu \nu}$.

In our absolute space framework, the force free condition reduces to

\begin{equation}
E \cdot J = 0,
\label{fofree1}
\end{equation}
and
\begin{equation}
\rho E + J \times B =0.
\label{fofree2}
\end{equation}

These restrictions, along with Maxwell's equations and Eqs.\ (\ref{suscept1}) and (\ref{suscept2}),
imply that

\begin{equation}
E_T = 0,
\label{etor}
\end{equation}
and
\begin{equation}
E_P \cdot B_P = 0.
\label{epol}
\end{equation}
The poloidal and toroidal components ($A_P$, and $A_T$) of a vector field are
defined such that $A = A_P + A_T$, where $A_P = A^r \partial_r + A^\theta \partial_\theta$
and $A_T = A^\varphi \partial_\varphi$. Eqs.\ (\ref{etor}) and (\ref{epol}) imply
that there exists a vector $\omega = \Omega \partial_\varphi$ such that

\begin{equation}
E = -\, \omega \times B,
\label{edef}
\end{equation}
From the vanishing of the curl of $E$
under the stationarity condition (Eq.\ (\ref{max2})), one finds that

\begin{equation}
B \cdot \nabla \Omega =0.
\label{bconstraint}
\end{equation}
It can also be shown that
\begin{equation}
B \cdot \nabla H_\varphi =0.
\label{hphispol}
\end{equation}

\section{Explicit Expressions for Fields and Currents}

To simplify calculations, we shall assume that the spatial
coordinates are orthogonal, and that the shift vector $\beta$
is purely toroidal, i.e., $\beta = (0,0,\beta_\varphi)$. 
The Kerr solution written in Boyer-Lindquist (though not
Kerr-Schild) coordinates can be written in this form.
%These assumptions are not very restrictive as they might appear, for
%even the Kerr Solution can be written in this form.

Surfaces of constant $\Omega$ are referred to as poloidal surfaces
(not to be confused with poloidal components of a vector). 
From Eq.\ (\ref{bconstraint}) it is clear that $B$ is tangent to poloidal
surfaces. Since $\Omega$ does not have any $\varphi$ dependence, and since
Eq.\ (\ref{bconstraint}) has nothing to say about the toroidal component of
$B$, it is clear that $B_P$ will entertain solutions of the type

\begin{equation}
B_P = \frac{\Lambda}{\sqrt\gamma}(-\Omega_{,\theta}  \partial_r + \Omega_{,r}
 \partial_\theta)
\label{bpexplicit}
\end{equation}
where, for the moment, $\Lambda$ is an arbitrary function. This must be so because
in the two dimensional subspace given by $\Omega = const$, there is a
unique vector (modulo magnitude) that is perpendicular to $\nabla \Omega$. The
condition that $B$ is divergence free means that $\Lambda$ satisfies

\begin{equation}
\Lambda_{,r}\, \Omega_{,\theta}\; = \; \Lambda_{,\theta}\, \Omega_{,r}.
\label{lambdadef}
\end{equation}
Consequently, $\Lambda$ is a poloidal function (a function that is constant on
poloidal surfaces). In the notation of the original paper by \citet{bz77},
$ \Lambda d\Omega \equiv -dA_\varphi$. The electric field is immediately calculated from
Eq.\ (\ref{edef}) and, as expected, is the gradient of a scalar function:
\begin{equation}
E_P = \Lambda d(\Omega^2/2) = d\int \Lambda \Omega d\Omega\,.
\label{eformula}
\end{equation}
From Eq.\ (\ref{suscept1}), we see that
\begin{equation}
D= D_P = \frac{\Lambda}{\alpha}(\Omega + \beta^\varphi) d\Omega.
\label{dformula}
\end{equation}
Similarly,  the expression for $H_P$ can be calculated from Eq.\ (\ref{suscept2}),
giving

\begin{equation}
H_P = (\alpha^2-\beta^2-\beta_\varphi \Omega)\frac{B_P}{\alpha}.
\label{hpolformula}
\end{equation}
The electric charge is determined by the divergence of the $D_P$,
(Eq.\ (\ref{max3})). Explicitly,

$$\sqrt \gamma \rho = \partial_r[\invdetst (\gamma_{\varphi \varphi} \Omega +\beta_\varphi)
\gamma_{\theta \theta} \Omega_{,r}]+$$
\begin{equation}
 \partial_\theta
[\invdetst (\gamma_{\varphi \varphi}\Omega +\beta_\varphi)\gamma_{r r} \Omega_{,\theta}].
\label{rhoformula}
\end{equation}
The toroidal component of the electric current density vector can be obtained
from the derivatives of components of $H_P$:

$$\sqrt \gamma J^\varphi = H_{\theta,r}-H_{r,\theta}=
\partial_r[\invdetst (\alpha^2-\beta^2-\beta_\varphi \Omega)
\gamma_{\theta \theta} \Omega_{,r}]+$$
\begin{equation}
\partial_\theta
[\invdetst (\alpha^2-\beta^2-\beta_\varphi \Omega)\gamma_{r r} \Omega_{,\theta}].
\label{jphiformula}
\end{equation}
It clear from the above discussion that the poloidal fields and, consequently,
the toroidal current $J^\varphi$ are uniquely described by the poloidal functions $\Omega$
and $\Lambda$.
On the other hand, the toroidal fields and the poloidal currents can be determined
from the poloidal function $H_\varphi$. In particular, from Eq.\ (\ref{suscept2}), it is
clear that $ H_\varphi = \alpha B_\varphi$. Maxwell's equation (Eq.\ (\ref{max4})) implies that

\begin{equation}
\sqrt \gamma J_p = H_{\varphi,\theta} \partial_r -H_{\varphi,r} \partial_\theta.
\label{jpolformula}
\end{equation}
Thus we see that fields and currents separate into two distinct categories: objects
that are determined by $\Omega$ and $\Lambda$, and those that are determined by $H_\varphi$.
Outside of the fact that $H_\varphi$ is a poloidal function (by definition, $\Omega$ is),
it is not yet clear as to how these two functions are dynamically related. This issue will
be cleared up in the following section.

\section {The Constraint Equation}

The expressions for the fields and currents given in the previous section naturally satisfies
Eq.\ (\ref{fofree1}).  Since the toroidal component of the electric field vanishes,
it is easily checked that, from Eq.\ (\ref{fofree2}), $(J \times B)_\varphi=0$ (as shown below in
Eq.\ (\ref{fofreefofree})). Thus the only
remaining requirements for a force-free solution is

\begin{equation}
\rho E_P + (J \times B)_P =0.
\label{fofreepol}
\end{equation}
The implication of the above equation is most easily understood by projecting the equation onto 
${E_P,B_p}$, which serve as a basis vectors for poloidal vector fields. 
The above equation yields no constraint when projected onto $B_P$, i.e.,

$$\rho E \cdot B_p + (J \times B) \cdot B_P=(J \times B) \cdot (B-B_T)=$$
\begin{equation}
-B^\varphi (J \times B)_\varphi=-B^\varphi \Lambda \invdetsp (H_{\varphi,\theta}\Omega_{,r}-H_{\varphi,r}\Omega_{,\theta})=0.
\label{fofreefofree}
\end{equation}
Projecting Eq.\ (\ref{fofree2}) onto $E_P$ gives

$$\rho E \cdot E_p + (J \times B) \cdot E_P=$$
$$\rho E^2+((J_P+J_T) \times (B_P+B_T)) \cdot E_P =$$
\begin{equation}
\rho E^2+((J_P \times B_T)+(J_T\times B_P)) \cdot E_P =0,
\label{finalconsa}
\end{equation}
since $J_P$ is parallel to $B_P$. With the help of the following relations,

$$J_P=\invdetsp \frac{d H_\varphi}{d \Omega}(\Omega_{,\theta} \partial_r-
\Omega_{,r} \partial_\theta)=-\frac{d H_\varphi}{\Lambda d \Omega}B_P,$$
$$E_r B_\theta-E_\theta B_r=\frac{\Omega \Lambda^2}{\sqrt\gamma}(\gamma_{\theta\theta} (\Omega_{,r})^2+\gamma_{r r} (\Omega_{,\theta})^2),$$
and
\begin{equation}
E^2=\frac{\Omega^2 \Lambda^2}{\gamma_{rr} \gamma_{\theta \theta}}(\gamma_{\theta\theta} (\Omega_{,r})^2+\gamma_{r r} (\Omega_{,\theta})^2),
\label{conshelp}
\end{equation}
Eq.\ (\ref{finalconsa}) reduces to the manageable form
\begin{equation}
\frac{1}{2 \Lambda } \frac{d H_\varphi^2}{d \Omega}=\alpha (\rho \Omega \gamma_{\varphi \varphi}-J_\varphi).
\label{finalconsb}
\end{equation}
This is the final and only constraint equation. If $\Omega$ and $\Lambda$ are picked such that the right hand side of the above equation is a poloidal function, then $H_\varphi$ continues to be poloidal function. The poloidal functions $\Omega$, $\Lambda$, and $H_\varphi$ then uniquely determines all currents and fields. It is important to realize that $\Omega$ is not to be thought of as a potential: physically relevant quantities like the electric field depend on $\Omega$ directly, and are not invariant transformations of the type $\Omega \rightarrow \Omega + const$. The charge density $\rho$ and the toroidal current $J_\varphi$ are functions of $\Omega$ and $\Lambda$ (see Eqs.\ (\ref{rhoformula}) and (\ref{jphiformula})). 
%The above equation can be integrated to obtain the result
% \begin{equation}
%H^2_\varphi = \pm H^2_0 + 2 \int \Lambda \alpha (\rho \Omega \gamma_{\varphi \varphi}-J_\varphi)d\Omega.
%\label{H0}
%\end{equation}

\section {Asymptotic Solutions and Energy Extraction}

Inserting Eqs.\ (\ref{rhoformula}) and (\ref{jphiformula}) into Eq.\ (\ref{finalconsb}), our
constraint equation gives
$$\frac{1}{2 \Lambda } \frac{d H_\varphi^2}{d \Omega}=\frac{\alpha \gamma_{\varphi \varphi}}{\sqrt\gamma}[\Omega \partial_r(\invdetst (\gamma_{\varphi\varphi} \Omega+ \beta_\varphi) \gamma_{\theta\theta} \Omega_{,r})+$$
$$\Omega \partial_\theta(\invdetst (\gamma_{\varphi\varphi} \Omega+ \beta_\varphi) \gamma_{rr} \Omega_{,\theta})+$$
$$\partial_r(\invdetst (\beta^2 - \alpha^2+\beta_\varphi \Omega) \gamma_{\theta\theta} \Omega_{,r})+$$
\begin{equation}
\partial_\theta(\invdetst (\beta^2 - \alpha^2+\beta_\varphi \Omega)\gamma_{rr} \Omega_{,\theta})].
\label{finalconsc}
\end{equation}
Therefore, Eq.\ (\ref{finalconsb}) is equivalent to Eq.\ (3.14) of \citet{bz77} written in the 3+1 formalism.

While searching for solutions for the fields and currents that might permit extraction of energy and angular momentum from a rotating black hole, it is advantageous to observe that
\begin{equation}
\frac{d^2 \cal E}{dA dt}=S^r \sqrt {\gamma_{rr}}=-H_\varphi \Omega B^r \sqrt {\gamma_{rr}},
\label{engext}
\end{equation}
\begin{equation}
\frac{d^2 \cal L}{dA dt}=L^r \sqrt {\gamma_{rr}}=-H_\varphi  B^r \sqrt {\gamma_{rr}},
\label{angext}
\end{equation}
as can be see from Eqs.\ (\ref{engflux}) and (\ref{angflux}). Here $\cal E$ and $\cal L$ are the total energy and angular momentum, respectively, extracted from the black hole.

For definiteness, we shall consider the magnetosphere 
of a Kerr black hole in Boyer-Lindquist coordinates. 
For finite rates of energy and angular momentum extraction, 
it is clear from the above two equations
that for $r \gg M$, $\Omega \rightarrow \Omega (\theta)$. With this in mind, we seek solutions to the constraint equation of the type $\Omega = \Omega (\theta)$ for all values of $r$. This means that all poloidal functions are functions of $\theta$ alone, since all poloidal functions are of ``zeroeth" order in $r$. Due to the inherent complexity of the constraint equation, we shall further consider Eq.\ (\ref{finalconsc}) in the far field limit. To order $(1/r^3)$ for strictly $\theta$-dependent functions, Eq.\ (\ref{finalconsc}) takes the form:

$$-\frac{1}{2 f(\theta)} \frac{d H_\varphi^2}{d \theta}=-\Omega \sin\theta \frac{d}{d\theta}(f\Omega \sin\theta)$$
$$+\frac{\sin\theta}{r^2}[-a^2\Omega\sin^2\theta\frac{d}{d\theta}(f\Omega \sin\theta)+\frac{d}{d\theta}(\frac{f}{\sin\theta})] $$
$$+2M\frac{\sin\theta}{r^3}[a\Omega\frac{d}{d\theta}(f \sin\theta(1-a\Omega\sin^2\theta))-$$
\begin{equation}
\frac{d}{d\theta}(\frac{f}{\sin\theta}(1-a\Omega\sin^2\theta))],
\label{assym}
\end{equation}
where $M$ and $a$ are the mass and the angular momentum per unit mass of the black hole, respectively, and $f(\theta)\equiv -\Lambda \Omega_{,\theta}\equiv A_{\varphi,\theta}$. For a consistent formulation of the theory of 
axisymmetric, stationary, force-free magnetospheres, the above equation implies that if $H_\varphi$ is to be a poloidal function of $\theta$ alone, then the terms proportional to the inverse powers of $r$ must vanish identically for choices of $f$ and $\Omega$.

General solutions to order $1/r^2$ can be considered by ignoring the $1/r^3$ term in the right hand side of Eq.\ (\ref{assym}). Here we require that $\Omega$ and $f$ satisfy the relation
\begin{equation}
a^2\Omega\sin^2\theta\frac{d}{d\theta}(f\Omega \sin\theta)=\frac{d}{d\theta}(\frac{f}{\sin\theta}).
\label{secondorder}
\end{equation}
To solve this, let $g \equiv f\Omega \sin\theta$ and $h \equiv (\Omega \sin^2\theta)^{-1}$. With these definitions,
Eq.\ (\ref{secondorder}) becomes 

\begin{equation}
\frac{a^2}{h} \frac{d}{d\theta} g = \frac{d}{d\theta} (gh).
\end{equation}
Integrating the above equation results in the relation

\begin{equation} 
g = \frac{C_1\Omega \sin^2\theta}{\sqrt{\mid a^2-h^2 \mid }}.
\end{equation}
Consequently, for an arbitrarily chosen $\Omega$,  

\begin{equation}
f = \frac{C_1 \sin\theta}{\sqrt{\mid (a\Omega\sin^2\theta)^2 -1 \mid}}
\label{feq}
\end{equation}
would make the $1/r^2$ term in Eq.\ (\ref{assym}) vanish. 
We can therefore successfully obtain 
the following function for $H_\varphi$. Explicitly,

$$\frac{d H_\varphi^2}{d \theta}=2f\Omega \sin\theta \frac{d}{d\theta}(f\Omega \sin\theta)=\frac{d}{d\theta}(f\Omega \sin\theta)^2\Rightarrow $$
\begin{equation} 
H_\varphi^2= \pm H_0 ^2+(f\Omega \sin\theta)^2.
\label{hphi}
\end{equation}
The choice of $\Omega$ is determined by the Znajek regularity
 condition applied at the event horizon ($r_+=M+\sqrt{M^2-a^2}$) 
so as to make $B_\varphi$ finite in the well-behaved (even near the event horizon) 
Kerr-Schild coordinate system (see  \citet{zna77,kom04}). The Znajek
condition can be written as 
\begin{equation} 
H_\varphi=\frac{\sin^2\theta}{\alpha_+}(2 r_+ M\Omega-a)B^r=\frac{\sin\theta}{\rho_+ ^2} (2r_+ M \Omega-a)f,
\label{EHBC}
\end{equation}
where the subscript $+$ indicates that the relevant quantities are to be evaluated at the event horizon and $\rho_+^2 = r_+^2 + a^2 \cos^2\theta$. 
From Eqs.\ (\ref{hphi}) and (\ref{EHBC}), we see that

\begin{equation} 
\pm H_0 ^2=\frac{\sin^2\theta}{\rho_+ ^4} [(4r_+^2 M^2-\rho_+ ^4) \Omega^2-4r_+ M a\Omega+ a^2]f^2.
\label{nonzeroho}
\end{equation}
We shall consider the solution for $H_\varphi$ such that $H_0 ^2=0$ (it is easily seen that when $H_0 \neq 0$, the resulting solution does not permit a finite rate of energy extraction, and this type of situation will be dealt with in  Subsection 6.2). This is possible if and only if the quantity in the square brackets in the above equation vanishes identically. Solving the resulting quadratic equation for $\Omega$, we find two solutions, namely

\begin{equation} 
\Omega_+ = \frac{a}{2 M r_+ + \rho_+ ^2}, 
\end{equation}
and 
\begin{equation} 
\Omega_- = \frac{a}{2 M r_+ - \rho_+ ^2}
= \frac{1}{a \sin^2\theta}.
\label{omegaminus}
\end{equation}

From Eq.(\ref{feq}) and the definition of $f$, we see that the only non-vanishing poloidal component of the magnetic field is given by
\begin{equation} 
B^r_{\pm}=\invdetsp f=\invdetsp \frac {B_0 \sin\theta (2 M r_+ \pm \rho_+ ^2)}{\sqrt{\mid (a \sin\theta)^4-(2 M r_+ \pm \rho_+ ^2)^2\mid}},
\label{br}
\end{equation}
where we have relabeled $C_1$ as $B_0$. It is clear that $\Omega_-$ is an unphysical solution since $B^r$ as given above is undefined everywhere.

\subsection{The $\Omega_+$ Solution}

In this case, the non-vanishing components of the fields are

$$B^r=\invdetsp \frac{B_0 \sin \theta}{2 \rho_+}\frac{\sqrt{a \Omega_H}}{\Omega_+}$$
$$E_\theta = -\sqrt{\gamma} \Omega_+ B^r$$
\begin{equation}
\alpha B_\varphi = H_\varphi = -\sqrt{\gamma} \Omega_+ B^r\sin\theta,
\label{omonefields}
\end{equation}
where $\Omega_H=a/2Mr_+$ is the angular velocity of the event horizon.  When $a\ll M$

$$B^r_+\rightarrow \invdetsp B_0 \sin\theta\,,{\rm~and}$$
\begin{equation}
\Omega_+ \rightarrow \frac{a}{8 M^2}.
\end{equation}
This is precisely the \citet{bz77} monopole solution 
\citep{kom04}. Therefore, the solutions for the fields and currents 
corresponding to $\Omega = \Omega_+$ generalizes the 
Blandford-Znajek monopole solution to accommodate the case of a black hole for all values 
of $a^2 < M^2$.

A parallel approach to the study of the force-free magnetosphere has been developed via the Grad-Shafranov equation (see, e.g.,  Eq.\ (6.4) of \citet{mdt82}). 
In our notation, the Grad-Shafranov equation takes the form \citep{uzd05}
$$\nabla \cdot [\frac{\alpha \nabla \psi}{\gamma_{\varphi \varphi}}(1-\frac{(\Omega_+ +\beta^\varphi)^2\gamma_{\varphi \varphi}}{\alpha^2})]$$
\begin{equation}
+\frac{(\Omega_+ +\beta^\varphi)}{\alpha}\frac{d \Omega_+}{d \psi}(\nabla \psi)^2+ \frac{I}{\alpha \gamma_{\varphi \varphi}} \frac {d I}{d \psi}=0 .
\label{GS}
\end{equation}
Here, $I=H_\varphi$ and $\psi = A_\varphi$. By straightforward substitution and evaluation 
of the various terms in eq.\ (\ref{GS}), it is not difficult to see that our 
solution satisfies the Grad-Shafranov equation to order $1/r^2$. 

From Eqs.\ (\ref{omonefields}) and (\ref{engext}), 
the angular dependence of energy extraction can be calculated.  
In the limit 
$r \gg M$, the result is 
\begin{equation}
\frac{d^2 \cal E}{dA dt}\approx \frac {a \Omega_H}{r^2} (\frac{B_0}{2})^2 \frac{\sin^2\theta}{\rho_+^2}.
\end{equation}
From the above equation, it is clear that most of the energy extraction 
happens along the equatorial plane. The total rate of energy extraction can be obtained by integrating the above result,
giving
\begin{equation}
\frac{d \cal E}{dt}=\frac{\pi B_0^2}{a r_+}[ \arctan \frac{a}{r_+}- \frac{a}{2M}].
\end{equation}

In similar fashion, we see by integrating Eq.\ (\ref{angext}) that
\begin{equation}
\frac{d \cal L}{dt}=\frac{2 \pi }{3} B_0^2  \Omega_H +  \frac{1}{\Omega_H}\frac{d \cal E}{dt} .
\label{raterel}
\end{equation}
As a result of energy and angular momentum extraction 
from the black hole, 
the mass and the total angular momentum ($J=aM$) of the black hole changes by the amount 
\begin{equation}
\frac{\delta M}{\delta t}=-\frac{d \cal E}{d t}, {\rm~and~}
 \frac{\delta J}{\delta t}=-\frac{d \cal L}{d t},
\end {equation}
respectively. From Eq.\ (\ref{raterel}) and the above definitions, it clear that
\begin{equation}
\frac{\delta J}{\delta t}+\frac{2 \pi }{3} B_0^2  \Omega_H =  \frac{1}{\Omega_H}\frac{\delta M}{\delta t} .
\end{equation}
Therefore we get the familiar inequality \citep{chr70}
\begin{equation}
\frac{\delta J}{\delta t}\leq  \frac{1}{\Omega_H}\frac{\delta M}{\delta t},
\end{equation}
which ensures that the irreducible mass of the black hole is non-decreasing if the black hole evolves along a Kerr sequence in a reversible way. This process therefore cannot  lead to the formation of a naked singularity. 

\subsection{A Jet-Type Solution}

It is easily seen that $\Omega = \Omega_-$ removes all the 
$r$-dependence in the right-hand side of Eq.\ (\ref{assym}) to order $1/r^3$. 
As shown by  Eqs.\  (\ref{omegaminus}) and (\ref{br}),
$\Omega = \Omega_-$ is not a physical solution for the condition 
$H_0 = 0$.
We now let $H_0 \neq 0$, and impose the 
Znajek  condition, Eq.\ (\ref{EHBC}), for this case.
Because our solutions involve both $\Omega_+$ and $\Omega_-$, 
the results continue to be valid only to order $r^{-2}$.

From Eq.(\ref{nonzeroho}) we see that
\begin{equation}
f^2 = \frac{\pm H_0^2 \rho_+^4}{a}\frac{\Omega_+ \Omega_-^2}{(\Omega-\Omega_+)(\Omega-\Omega_-)}.
\end{equation}
Here the $\pm$ factor is to ensure that $f^2 \geq 0$. Similarly we find from Eq.\ (\ref{feq}) that 
\begin{equation}
f^2 = \frac{B_0^2}{a^2 \sin^2\theta \mid(\Omega-\Omega_-)(\Omega+\Omega_-)\mid }.
\label{fp}
\end{equation}

Equating the right-hand sides of the last two equations, we see that
\begin{equation}
B_0^2  \mid \Omega-\Omega_+\mid =\frac{\pm H_0^2 \rho_+^4}{ \sin^2\theta(2Mr_+ +\rho_+^2)}
\mid \Omega+\Omega_- \mid.
\end{equation}
It is important to remember that any $\Omega$ satisfying the above equation 
is consistent with Eq.\ (\ref{assym}) (to order $1/r^2$) 
and with Eq.\ (\ref{EHBC}). The above equation has the unique solution
\begin{equation}
\Omega_p =\frac{\tilde{A}\Omega_+ +\tilde{B}\Omega_-}{\tilde{A}-\tilde{B} }\,,
\end{equation}
where

$$\frac{\tilde{A}}{B_0^2} = \cases {+ 1, &if $\Omega_p - \Omega_+ \geq 0$\cr-1, &otherwise\cr}\,,{\rm ~and}$$
\begin{equation}
\frac{\tilde{B}}{H_0^2 \rho_+^4 \Omega_+ \Omega_-
}= \cases {+ 1, &if $\Omega_p + \Omega_- \geq 0$\cr-1
, &otherwise.\cr}
\label{constcondition}
\end{equation}
All other poloidal fields quantities are now uniquely determined by noting that $f$ is given by Eq.\ (\ref{fp}).
It is important to see if we can indeed satisfy the above conditions. A quick calculation shows that

$$\Omega_p - \Omega_+ = \frac{\tilde{B}(\Omega_+ +\Omega_-)}{\tilde{A}-\tilde{B}}\;,{\rm~and}$$
\begin{equation}
\Omega_p + \Omega_- = \frac{\tilde{A}(\Omega_+ +\Omega_-)}{\tilde{A}-\tilde{B}}.
\end{equation}
Therefore the choice $\tilde{A}=-B_0^2$ and $\tilde{B}=+ H_0^2 \rho_+^4 \Omega_+ \Omega_-
 $ is a valid one. We shall pick this choice for the remainder of the paper. Consequently, we have
\begin{equation}
\Omega_p = -\Omega_+ \Omega_- \frac{[H_0^2\rho_+^4-B_0^2 a^2 \sin^4\theta]}{[H_0^2\rho_+^4\Omega_+ +B_0^2 a \sin^2\theta]}.
\end{equation}
Note that as $\theta \rightarrow 0$ and $\pi$, $~\Omega_p \rightarrow -\Omega_- \rightarrow -\infty $ .
The form of $f$ is determined by Eq.\ (\ref{fp}), and upon substitution of the explicit form of $\Omega$ ($=\Omega_p$), we find that in the limit as $\theta \rightarrow 0$ and $\pi$, $ f \rightarrow \pm H_0  \sqrt{a/\Omega_H}/2$.  

The expression for the rate of total energy extraction is given by
\begin{equation}
\frac{d^2 \cal E}{dA dt}= -H_\varphi \Omega_p \invdetsp f \approx \frac {- 1}{r^2 }f^2 \Omega_p \frac{(2 M r_+ \Omega_p -a)}{\rho_+^2}.
\end{equation}
As $\theta \rightarrow 0$ and $\pi $,
\begin{equation}
\frac{d^2 \cal E}{dA dt} \rightarrow \frac{-1}{r^2} \frac{H_0^2 a}{4 \Omega_H} \Omega_-(\Omega_- + \Omega_H)
\end{equation}
A solution of this type has the following features:
 Energy extraction is less than zero near the poles, i.e., energy is being fed into the system, indicating a reverse jet type situation.
Also, the total rate of energy and angular momentum ``insertion" is not calculable since the above integral is divergent along the poles. This solution is therefore unphysical.

\section {Discussion}
Based on the 3+1 equations as written by \citet{kom04}, we have
rederived the constraint equation relating the toroidal magnetic field
to the charge and current densities in a force-free magnetosphere around
a spinning black hole.
Known solutions to the constraint equation for the force-free
magnetosphere include the  monopole and the parabolic solutions obtained
in the orginal paper by \citet{bz77}, and the solution by \citet{bip92} 
for a black hole 
surrounded by a magnetized, conducting accretion disk. 
We have discovered a solution to the
constraint equation  that  generalizes the ``monopole solution" originally derived by
\citet{bz77}. This solution satisifes the \citet{zna77} 
regularity condition at the event horizon, even in the limit
$a/M \ll 1$ (contrary to the statement of Blandford and Znajek).

\citet{kom01} has used a time-dependent
numerical simulation to calculate the electromagnetic
extraction of energy for a monopole magnetic
field at different values of a/M. Our value
of $\Omega_+/\Omega_H$ at $\theta = 0.5$
ranges from 0.5 to 0.58 when $a/M$ varies
from $0.1$ to $0.9$, in comparison with the
numerical value of $0.52$ for $a/M = 0.9$
at $r = 10$.  The value of $H_\varphi$ for our
$\Omega_+$ solution is $\approx 25$\% larger
than the numerical value of \citet{kom01} 
when $a/M = 0.9$. These discrepancies, though
not large, may reflect the finite
value of $r = 10$ used in Komissarov's work, whereas
our solution holds in the asymptotic limit of large
$r$.

For the $\Omega_+$ solution, energy and angular momentum is extracted preferentially
along the equatorial directions of the spinning black hole. As such, it does
not account for galactic black holes
and active galactic nuclei that display radio jets. Time-dependent 
numerical solutions employing accretion of magnetized plasma into the ergosphere
seem to indicate the presence  of such jet-like features \citep{sdp04}.  
In future work, analytic solutions that
exhibit jet-like structures will be studied using the techniques developed in
this paper.

\vskip0.5in
We acknowledge funding for this research
through NASA {\it GLAST} Science Investigation No.\ DPR-S-1563-Y.
The work of C.\ D.\ D.\ is supported by the Office of Naval Research. The authors wish to thank the referee for providing valuable comments.

\clearpage


\begin{thebibliography}{}

\bibitem[Beskin et al.(1992)]{bip92} Beskin, V.~S., Istomin, Y.~N., 
\& Pariev, V.~I.\ 1994, in Extragalactic Radio Sources.~From Beams to Jets, 
ed.\ Roland, J., Sol, H., \&  Pelletier, G.\ (Cambridge U.\ Press), 45

\bibitem[Blandford \& Znajek(1977)]{bz77} Blandford, R.~D., 
\& Znajek, R.~L.\ 1977, \mnras, 179, 433 

\bibitem[Christodoulou(1970)]{chr70} Christodoulou, D.\ 1970, 
Physical Review Letters, 25, 1596 

\bibitem[Komissarov(2001)]{kom01} Komissarov, S.~S.\ 2001, 
\mnras, 326, L41 

\bibitem[Komissarov(2004)]{kom04} Komissarov, S.~S.\ 2004, 
\mnras, 350, 427 


\bibitem[MacDonald \& Thorne(1982)]{mdt82} MacDonald, D., \& 
Thorne, K.~S.\ 1982, \mnras, 198, 345 

\bibitem[Penrose(1969)]{pen69} Penrose, R.\ 1969,  Nuovo Cimento, 1, 252 

\bibitem[Punsly \& Coroniti(1990)]{pc90} Punsly, B., \& 
Coroniti, F.~V.\ 1990, \apj, 350, 518 

\bibitem[Punsly(2001)]{pun01} Punsly, B.\ 2001, Black Hole 
Gravitohydromagnetics (Berlin: Springer)

%\bibitem[Roland et al.(1992)]{rol92} Roland, J., Sol, H., \& 
%Pelletier, G.\ 1992, Extragalactic Radio Sources.~From Beams to Jets


\bibitem[Semenov et al.(2004)]{sdp04} Semenov, V., 
Dyadechkin, S., \& Punsly, B.\ 2004, Science, 305, 978 


\bibitem[Thorne \& MacDonald(1982)]{tmd82} Thorne, K.~S., \& 
MacDonald, D.\ 1982, \mnras, 198, 339 

\bibitem[Thorne et al.(1986)]{tpm86} Thorne, K.~S., Price, 
R.~H., \& MacDonald, D.~A.\ 1986, Black Holes: The Membrane Paradigm,  (New Haven: Yale)

\bibitem[Uzdensky(2005)]{uzd05} Uzdensky, D.~A.\ 2005, \apj, 
620, 889 

 
\bibitem[Znajek(1977)]{zna77} Znajek, R.~L.\ 1977, \mnras, 
179, 457 

\end{thebibliography}
\end{document}